\documentclass[aps,prl,twocolumn,showpacs,floatfix]{revtex4} 
\usepackage{amsmath,amssymb,isolatin1,graphicx,times}

\newcommand{\be}{\begin{equation}}
\newcommand{\ee}{\end{equation}}
\newcommand{\bea}{\begin{eqnarray}}
\newcommand{\eea}{\end{eqnarray}}
\newcommand{\bw}{\begin{widetext}}
\newcommand{\ew}{\end{widetext}}

\newcommand{\kommentar}[1]{}

\begin{document}
 
\title{Universal Behavior of Quantum Walks with Long-Range Steps}
\author{Oliver M{\"u}lken}
\email{muelken@physik.uni-freiburg.de}
\author{Volker Pernice}
\author{Alexander Blumen}
\affiliation{
Theoretische Polymerphysik, Universit\"at Freiburg,
Hermann-Herder-Stra{\ss}e 3, 79104 Freiburg, Germany}

\date{\today} 
\begin{abstract}
Quantum walks with long-range steps $R^{-\gamma}$ ($R$ being the distance
between sites) on a discrete line behave in similar ways for all
$\gamma\geq2$. This is in contrast to classical random walks, which for
$\gamma >3$ belong to a different universality class than for $\gamma \leq
3$. We show that the average probabilities to be at the initial
site after time $t$ as well as the mean square displacements are of the
same functional form for quantum walks with $\gamma=2$, $4$, and with
nearest neighbor steps. We interpolate this result to arbitrary
$\gamma\geq2$.
\end{abstract}
\maketitle


One-dimensional models are not only prime toys for theoretical physicists
but also allow for deep physical insights. For instance, in solid state
physics lattice models describe the behavior of metals quite acurately
\cite{Ziman,Ashcroft}. Over the years these models have been refined and
augmented to address different phenomena, such as the dynamics of atoms in
optical lattices and the Anderson localization in systems with energetic
disorder \cite{anderson1958}. Classical one-dimensional models allow to
address various aspects of normal and anomalous diffusion
\cite{metzler2000}. 

The simplest model describing a particle moving on a regular structure
assumes only jumps from one position $j$ to its nearest neighbors (NN)
$j\pm1$. The tight-binding approximation for such systems is equivalent to
the so-called continuous-time quantum walks (CTQW), which model quantum
dynamics of excitations on networks \cite{farhi1998,mb2005a,bose2003}. Recently,
there have been several experimental proposals addressing CTQW in various
types of systems, ranging from microwave cavities \cite{sanders2003},
waveguide arrays \cite{hagai2007}, atoms in optical lattices
\cite{duer2002,cote2006}, or structured clouds of Rydberg atoms
\cite{mbagrw2007}. A large class of these systems do not show NN steps.
Consider, for instance, a chain of clouds of Rydberg atoms where each
cloud can contain only one excitable atom due to the dipole blockade
\cite{anderson1998,mbagrw2007}. The excited atoms of different clouds
interact via long-range couplings decaying as $R^{-3}$, where $R$ is the
distance between different clouds.   

The dynamics of classical excitations can be efficiently described by
continuous-time random walks \cite{Weiss}. Here, is has been shown that
CTRW in one dimension with step lengths decaying as $R^{-\gamma}$ belong
only to the same universality class if $\gamma>3$.  Those CTRW show normal
diffusion, whereas CTRW with $\gamma<3$ show anomalous diffusion as, e.g.,
L{\'e}vy flights. The reason is that the for $\gamma<3$ the second moment
of the step-length distribution $\langle R^2\rangle$ diverges
\cite{mbagrw2007,klafter1987}.

In the following we will consider in one dimension the dependence of the
dynamics of excitations on the range of the step length. We restrict
ourselves to the extensive cases, i.e., we explicitly exclude ultra-long
range interactions, where the exponent $\gamma$ of the decay of the step
length is smaller than the dimension ($\gamma<d$, $\gamma=d$ is the
marginal case); thus we take here $\gamma\geq2$. The effect of ultra-long
range interactions on the thermodynamics and dynamics of regular
one-dimensional lattices has been studied numerically before
\cite{borland1999}.

Our analysis is based on the density of states (DOS) of the corresponding
Hamiltonian. The DOS contains the essential information about the system
and allows to calculate various dynamical quantities, such as the
probability to be at time $t$ at the initially excited site.

The coherent dynamics of excitons on a graph of connected nodes is modeled
by CTQW, which follows by identifying the Hamiltonian ${\bf H}$ of the
system with the CTRW transfer matrix ${\bf T}$, i.e., ${\bf H} = - {\bf
T}$; see e.g.\ \cite{farhi1998,mb2005a} (we will set $\hbar \equiv 1$ in
the following).  For NN step lengths and identical transfer rates, ${\bf
T}$ is related to the connectivity matrix ${\bf A}$ of the graph by ${\bf
T} = - {\bf A}$.  In the following, we will consider one-dimensional
networks with periodic boundary conditions (i.e., a discrete ring).  Here,
when the interactions go as $R^{-\gamma}$, with $R=|k-j|$ being the (on
the ring minimal) distance between two nodes $j$ and $k$, the Hamiltonian
has the following structure:
\be
{\bf H}_\gamma = \sum_{n=1}^{N} \sum_{R=1}^{R_{\rm max}}R^{-\gamma}\Big(2 | n
\rangle \langle n| - | n-R \rangle
\langle n | - | n+R \rangle \langle n |\Big) ,
\ee
where $R_{\rm max}$ is a cut-off for finite systems. Note, that in the
infinite system limit we first take $N\to\infty$ before taking also
$R_{\rm max}\to\infty$.  For the cases considered here, namely
$\gamma\geq2$ and $N$ of the order of a few hundred nodes, a resonable
cut-off is $R_{\rm max}=N/2$, which is also the largest distance between
two nodes on the discrete ring. In this way, to each pair of sites a
single (minimal) distance and a unique interaction is assigned.

The states $|j\rangle$ associated with excitons localized at the nodes $j$
($j=1,\dots,N$) form a complete, orthonormal basis set (COBS) of the whole
accessible Hilbert space, i.e., $\langle k | j \rangle = \delta_{kj}$ and
$\sum_k |k~\rangle\langle~k| = {\bf 1}$. In general, the transition
amplitudes from state $|j\rangle$ to state $|k\rangle$ during $t$ and the
corresponding probabilities read $\alpha_{kj}^{(\gamma)}(t) \equiv \langle
k | \exp(-i {\bf H}_\gamma t) | j \rangle$ and $\pi_{kj}^{(\gamma)}(t)
\equiv \big| \alpha_{kj}^{(\gamma)}(t) \big|^2$, respectively. In the
classical CTRW case the transition probabilities obey a master equation
and can be expressed as $p_{kj}^{(\gamma)}(t) = \langle k | \exp({\bf
T}_\gamma t) | j \rangle$ \cite{farhi1998,mb2005a}.


For all $\gamma$, the time independent Schr\"odinger equation ${\bf
H}_\gamma | \Phi_\theta \rangle = E_\gamma(\theta) | \Phi_\theta \rangle$
is diagonalized by Bloch states $|\Phi_\theta\rangle =
N^{-1/2}\sum_{j=1}^{N} \exp(i \theta j) |j\rangle$.  One obtains the
eigenvalues 
\be
E_\gamma(\theta) = \sum_{R=1}^{R_{\rm max}} R^{-\gamma} \big[ 2 -
2\cos(\theta
R)\big].
\label{evals}
\ee

In the limit $N\to\infty$ the $\theta$-values are quasi-continuous. Then,
the density of states (DOS) $\rho_\gamma(E)$ is obtained by inverting
Eq.~(\ref{evals}) and taking the derivative with respect to $E_\gamma$.
In the NN-case ($\gamma=\infty$) only the first term in Eq.~(\ref{evals}),
$R=1$, contributes. From this we get the known DOS $\rho_\infty(E) =
\big(\pi\sqrt{4E-E^2}\big)^{-1}$.  For $\gamma=2$ we can approximate the
sum by letting $R_{\rm max} \to \infty$, which yields $E_2(\theta) =
\pi\theta - \theta^2/2$ (see Eq.~1.443.3 of \cite{gradshteyn}).  By
inverting this and assuming $\theta$ to be continuous one obtains
$\rho_2(E) = \big( \pi\sqrt{2}\sqrt{\pi^2/2 - E}\big)^{-1}$.  In the
intermediate range we have an analytic solution for $\gamma=4$, namely we
have $E_4(\theta) = \theta^4/24 - \pi\theta^3/6 + \pi^2\theta^2/6$ (see
Eq.~1.443.6 of \cite{gradshteyn}), which yields \cite{abramowitz}
$\rho_4(E) = \Big[2\pi (2/3)^{1/4}\sqrt{E (\pi^2/\sqrt{24}) - E^{3/2}}
\Big]^{-1}$.

In order to interpolate between $\rho_2(E)$ and $\rho_\infty(E)$ to
arbitrary values of $\gamma \in[2,\infty]$ we assume the following general
form for the DOS: 
\be
\rho_\gamma(E)
\sim \Big[ \sqrt{c_\gamma E^\alpha - E^\beta} \Big]^{-1}
\label{dos_gen}
\ee
with $\alpha\in[0,1]$ and $\beta\in[1,2]$; $c_\gamma$ is a constant
related to the maximal energy, $c_\gamma \equiv (E_{\gamma,{\rm
max}})^{\beta-\alpha}$.  Thus, for $\gamma=2$: $\alpha=0$ and $\beta=1$
($c_2=\pi^2/2$). for $\gamma=4$: $\alpha=1$ and $\beta=3/2$
($c_4=\pi^2/\sqrt{24}$), and for NN-walks: $\alpha=1$ and $\beta=2$
($c_\infty=4$).  Note that close to the band edge $\theta=0$, i.e., for
small $E$, Eq.~(\ref{dos_gen}) can be approximated by
\be
\rho_\gamma(E) \sim \begin{cases}
E^{-1/2} & \alpha=1 \ (\gamma>3) \\
E^{-\alpha/2} & \alpha<1 \ (2\leq\gamma\leq3) \end{cases},
\label{dosapprox}
\ee
from which we observe the distinction between $\gamma$-values larger and
smaller than three. For the band edge $\theta=\pi$, i.e., $E\approx E_{\rm
max}$, it is straightforward to show that $\rho(E) \sim (E_{\rm
max}-E)^{-1/2}$ for all $\gamma\geq2$. The behavior of
Eq.~(\ref{dosapprox}) for small $E$ is in line with previous studies
\cite{rodriguez2003}, in which the DOS goes as $\rho(E) \sim E^{\nu}$,
where $\nu=-1/2$ for $\gamma>3$ and $\nu=-(\gamma-2)/(\gamma-1)$ for
$2\leq\gamma<3$.  Starting from the two limiting cases $\gamma=2$ and
$\gamma=\infty$ and supported by the $\gamma=4$ case, Eq.~(\ref{dos_gen})
appears as a natural candidate for a generalized DOS. 

\begin{figure}[htb]
\centerline{\includegraphics[clip=,width=\columnwidth]{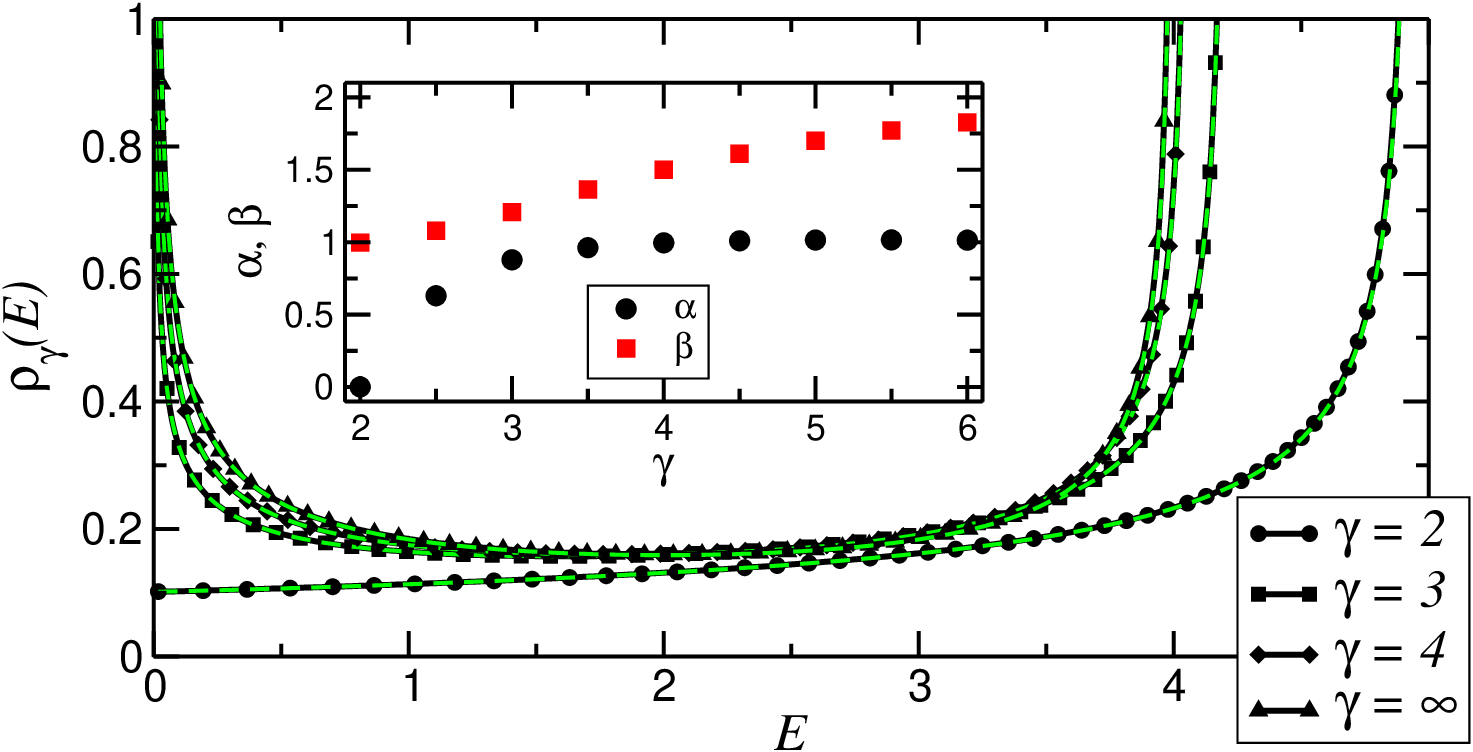}}
\caption{(Color online) Density of states for a discrete ring with
$N=10000$ nodes with hopping parameters $\gamma=2$, $3$, $4$, and $\infty$
(solid black curves with symbols), obtained by numerically diagonalizing
${\bf H}_\gamma$. The dashed green curves show the analytic expressions for
$\rho_2(E)$, $\rho_4(E)$, and $\rho_\infty(E)$ as well as the fit for
$\rho_3(E)$ [Eq.~(\ref{dos_gen})] given in the text. The inset shows the
exponents $\alpha$ and $\beta$ obtained by fitting the DOS for different
values of $\gamma$ to Eq.~(\ref{dos_gen}).
}
\label{dos}
\end{figure}

Figure~\ref{dos} shows a comparison of the DOS obtained from the numerical
diagonalization of ${\bf H}_\gamma$ for $N=10000$ with $\gamma=2$, $3$,
$4$, and $\infty$ (solid black curves) with the exact expressions for
$\rho_2(E)$, $\rho_4(E)$, and $\rho_\infty(E)$, see above, as well as a
fit for $\rho_3(E)$. The values of $\alpha$ and $\beta$, extracted from
fits to the numarical DOS for various values of $\gamma$, are given in the
inset of Fig.~\ref{dos}. Clearly, for $\gamma\geq4$ we have $\alpha=1$,
while $\beta\in]1,2[$. For $\gamma=2$, the values of $\alpha$ and $\beta$
drop to $\alpha=0$ and $\beta=1$, respectively.


CTRW with step widths distributed according to $R^\gamma$ belong to the
same universality class for $\gamma>3$, the mean square displacement (MSD)
going as $\langle R^2 \rangle \sim t$, i.e., showing normal diffusion, see
e.g.\ \cite{Weiss}.  For $\gamma\le3$ the second moment of the
distribution diverges, which leads to a MSD showing anomalous diffusion. 

Another way to see this is using the {\sl average} probability to be at
the initial site at time $t$, $\overline{p}_\gamma(t)$.  Classically one
has a simple expression for $\overline{p}_\gamma(t)$
\cite{alexander1981,bray1988},
\be
\overline{p}_\gamma(t) \equiv \frac{1}{N} \sum_{j=1}^{N}
p^{(\gamma)}_{j,j}(t)=
\frac{1}{N} \sum_{\theta} \ \exp[-E_\gamma(\theta) t],
\label{pclavg}
\ee
which depends only on the eigenvalues but {\sl not} on the eigenvectors.
In the quantum case, the corresponding expression is
$\overline{\pi}_\gamma (t) \equiv \frac{1}{N} \sum_{j=1}^{N}
\pi^{(\gamma)}_{j,j}(t)$.  For  the discrete ring, we get
\be
\overline{\pi}_\gamma(t) = |\overline{\alpha}_\gamma(t)|^2 =
\Big|\frac{1}{N} \sum_{\theta} \ \exp[-iE_\gamma(\theta)t] \Big|^2,
\label{pqmavg}
\ee
which also depends only on the eigenvalues. Note that for more complex
networks the right-hand-side of Eq.~(\ref{pqmavg}) is only a lower bound
to $\overline{\pi}_\gamma (t)$ \cite{mb2006b}.  In the continuum limit,
Eqs.~(\ref{pclavg}) and (\ref{pqmavg}) can be written as
\bea
\overline{p}_\gamma(t) &=& \int dE \ \rho_\gamma(E) \
\exp(-E t), 
\label{pclavginf} \\
\overline{\pi}_\gamma(t) &=& \Big| \int dE \ \rho_\gamma(E) \
\exp(-i E t) \Big|^2 .
\label{pqmavginf} 
\eea

Having the DOS at hand the integrals in Eqs.~(\ref{pclavginf}) and
(\ref{pqmavginf}) can be calculated - at least asymptotically - for large
$t$.  In the classical case Eq.~(\ref{pclavginf}) will be dominated by
small values of $E$ when $t$ becomes large, see Eq.~(\ref{dosapprox}).
From the DOS we obtain
\be
\overline{p}_\gamma(t) \sim \begin{cases} t^{-1/2} & \alpha=1 \\
t^{\alpha/2-1} & \alpha<1. \end{cases}
\ee

Quantum mechanically, some care is in order. Here, the assumption that
$\overline{\pi}_\infty(t)$ will be dominated by small values of $E$ for
large $t$ is not applicable, due to the oscillating exponential in
Eq.~(\ref{pqmavginf}). For the NN-case we know that
$\overline{\pi}_\infty(t) \sim t^{-1}$, see for instance \cite{mb2006b}.
Considering now the other limiting case $\gamma=2$ we have
\bea
\overline{\pi}_2(t) &=& \Big| \int_0^{\pi^2/2} dE \ \exp(-i E
t)\Big/\pi\sqrt{2}\sqrt{\pi^2/2 - E} \ \Big|^2 \nonumber \\
&=& \Big|\int_0^{\pi^2/2} d\epsilon \ \exp(-i \epsilon
t)\Big/\sqrt{2\pi^2\epsilon} \ \Big|^2 \sim t^{-1},
\label{pi2}
\eea
where we substituted $\epsilon \equiv \pi^2/2 - E$. Note that for $t\gg1$,
Eq.~(\ref{pi2}) approaches $\overline{\pi}_2(t) \approx (2\pi t)^{-1}$.
Thus, the dependence of $\overline{\pi}_2(t)$ on $t$ is the same as for
$\overline{\pi}_\infty(t)$. This suggests that for all one-dimensional
lattices with extensive ($\gamma\geq2$) interactions the long time
dynamics of the excitations is similar, no matter how long- or short-range
the step lengths are. This is in contrast to the classical case, where
only CTRW with $\gamma>3$ belong to the same universality class.

\begin{figure}[htb]
\centerline{\includegraphics[clip=,width=\columnwidth]{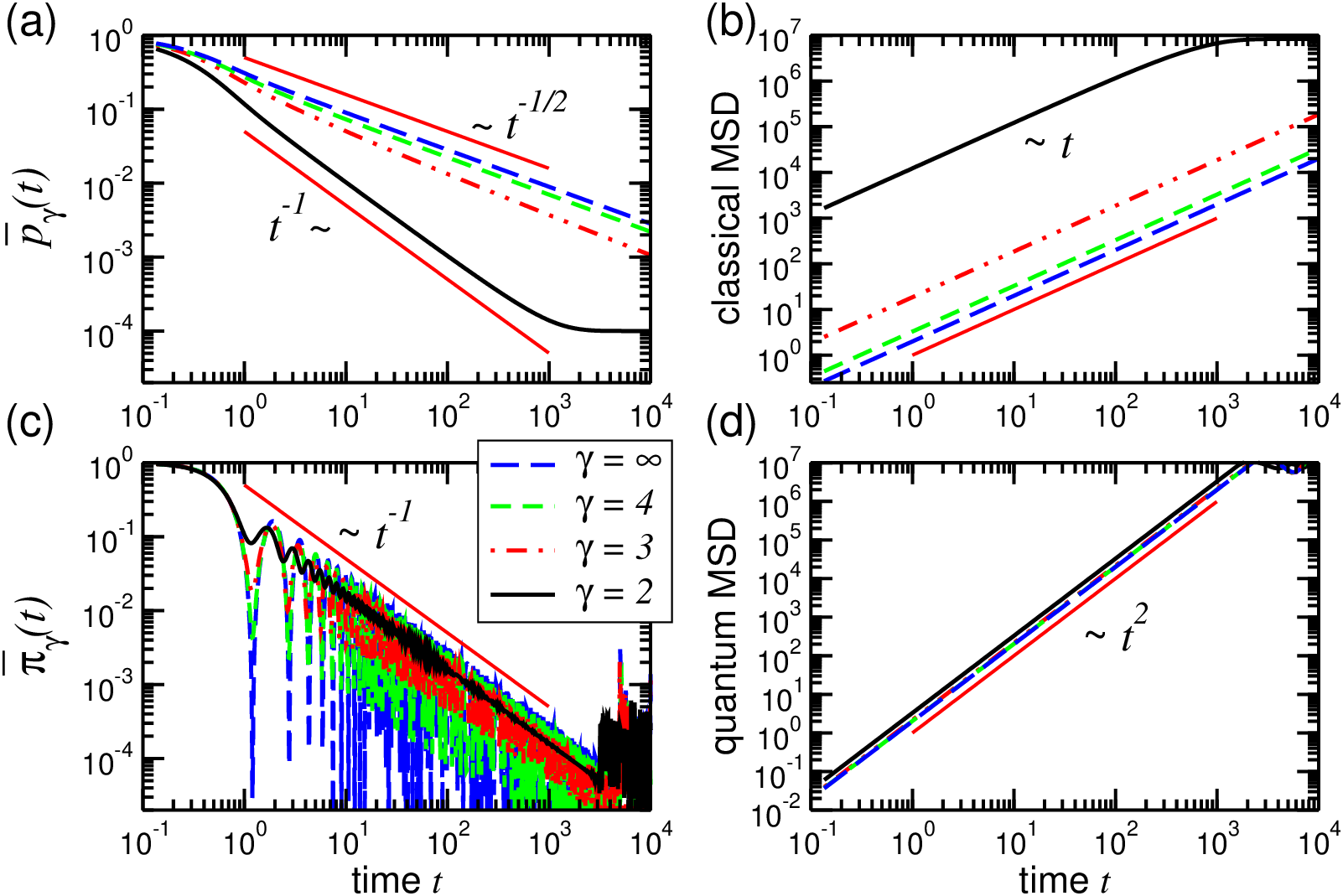}}
\caption{(Color online) (a) Classical $\overline{p}_\gamma(t)$ and (b)
corresponding MSD; (c) quantum mechanical
$\overline{\pi}_\gamma(t)$ and (d) corresponding MSD (right) for a
discrete ring with $N=10000$ nodes with $\gamma=2$, $3$,
$4$, and $\infty$. 
}
\label{returnprob}
\end{figure}

To test this we calculated numerically for a discrete ring of $N=10000$
nodes $\overline{p}_\gamma(t)$ and $\overline{\pi}_\gamma(t)$ for
different $\gamma$; the results are shown in Fig.~\ref{returnprob}.
Clearly, $\overline{p}_\gamma(t)$ changes when increasing the step width
from NN steps ($\gamma=\infty$) to long-range steps distributed as
$R^{-2}$ ($\gamma=2$), see Fig.~\ref{returnprob}(a).  While
$\overline{p}_\gamma(t)$ for $\gamma>3$ decays as $t^{-1/2}$, the power
law changes to $t^{-1}$ for $\gamma=2$. In contrast, the decay of the
maxima of the quantum return probability $\overline{\pi}_\gamma(t)$
follows $t^{-1}$ for all $\gamma$, Fig.~\ref{returnprob}(c).  Long-range
steps lead only to a damping of the oscillations and to an earlier
interference once the excitation has propagated half around the ring. 

The classical and quantum MSD corroborate these findings, see
Figs.~\ref{returnprob}(b) and \ref{returnprob}(d). Now, the MSD for
CTRW/CTQW on the discrete ring with initial site $j$ are given by
\be
\langle R_\gamma^2 (t) \rangle_{\rm cl;~qm} = \frac{1}{N}\sum_{k=1}^{N}
|k-j|^2 {\cal P}_{k,j}^{(\gamma)} (t),
\ee
where ${\cal P}_{k,j}^{(\gamma)} (t) = p_{k,j}^{(\gamma)} (t)$ for CTRW
and ${\cal P}_{k,j}^{(\gamma)} (t) = \pi_{k,j}^{(\gamma)} (t)$ for CTQW.
Now, decreasig $\gamma$ has huge effects on the classical MSD. For
$2\leq\gamma<3$ the MSD starts to diverge, which in the case of finite
networks is reflected in the fact that the MSD is of the order of $N$
already for very short times. Increasing $\gamma$ to values larger than
$3$ leads to the expected diffusive behavior $\langle R_\gamma^2 (t)
\rangle_{\rm cl} \sim t$ for all $\gamma>3$.  The quantum MSD, on the
other hand, do not diverge for all $\gamma$-values considered here. All
step lengths lead to the same qualitative behavior, $\langle R_\gamma^2
(t)\rangle_{\rm qm} \sim t^2$.  

Figure \ref{returnprob} also shows that the MSD can be related to
$\overline{p}_\gamma(t)$ and $\overline{\pi}_\gamma(t)$, through: 
\be
\langle R_\gamma^2 (t)
\rangle_{\rm cl;\ qm} \sim \begin{cases} [\overline{p}_\gamma(t)]^{-2} &
\\
[\overline{\pi}_\gamma(t)]^{-2} & 
\end{cases}
\label{msd}
\ee 
for $\gamma>3$ in the classical and $\gamma\geq2$ in the quantal case.
This generalizes previous (classical) results, obtained for regular
networks with NN-steps \cite{alexander1981}, to the quantum case and to
long-range steps.

We can now underline our results by analytically evaluating
$\overline{\pi}_\gamma(t)$ [Eq.~(\ref{pqmavg})] using the stationary phase
approximation (SPA) \cite{Bender}. We expect in general $E_\gamma(\theta)$
to be a smooth real-valued function on the interval $\theta\in[0,2\pi[$.
For large $N$, we  write $\overline{\alpha}_\gamma(t)$
[Eq.~(\ref{pqmavg})] in the integral form
\[
\overline{\alpha}_\gamma(t) = \frac{1}{2\pi}\int_0^{2\pi} d\theta \
\exp(iE_\gamma(\theta) t).
\]
The SPA asserts now that the main contribution to this integral comes from
those points where $E_\gamma(\theta)$ is stationary
[$dE_\gamma(\theta)/d\theta\equiv E_\gamma'(\theta)=0$]. If there is only
one point $\theta_0$ for which $E_\gamma'(\theta_0)=0$ and
$d^2E_\gamma(\theta)/d\theta^2|_{\theta_0}\equiv E_\gamma''(\theta_0)
\neq0$ one gets, see \cite{Bender},
\be
\overline{\alpha}_\gamma(t) \approx 
\frac{1}{\sqrt{2\pi t|E_\gamma''(\theta_0)|}} \ \exp\Big(i\Big\{tE_\gamma(\theta_0) +
\frac{\pi}{4}{\rm sgn}[E_\gamma''(\theta_0)]\Big\}\Big)
\label{spa1}
\ee
such that
\be
\overline{\pi}_\gamma(t) = |\overline{\alpha}_\gamma(t)|^2
\approx \frac{1}{2\pi t |E_\gamma''(\theta_0)|} \sim t^{-1}.
\ee 
For the infinite one-dimensional regular network [see Eq.(\ref{evals}),
where $N\to\infty$ and $R_{\rm max}\to\infty$] and for $\gamma=2$,
$E_2(\theta)$ (see above) has only one stationary point in
$\theta\in[0,2\pi[$, namely $\theta_0=\pi$. Then $E_2(\pi) = \pi^2/2$ and
$E_2''(\pi) = -1$, leading to $\overline{\pi}_2(t) \approx (2\pi t)^{-1}$,
which does not show any oscillations and coincides with the long time
limit of Eq.~(\ref{pi2}).  

For $\gamma>2$, $E_\gamma(\theta)$ [see Eq.~(\ref{evals})] has two
stationary points in the interval $\theta\in[0,2\pi[$, namely $\theta_0=0$
and $\theta_0=\pi$. Then $\overline{\alpha}_\gamma(t)$ is approximately
given by the sum of the contributions [each being of the form given in
Eq.~(\ref{spa1})] of the two stationary points. One easily verifies from
$E_\gamma''(\theta) = 2\sum_R \cos(\theta R)/R^{\gamma-2}$ that ${\rm
sgn}[E_\gamma''(0)]=1$ and ${\rm sgn}[E_\gamma''(\pi)]=-1$. Consequently,
we obtain   
\bea
\overline{\pi}_\gamma(t) &=& |\overline{\alpha}_\gamma(t)|^2 \approx \frac{1}{2\pi t} \Bigg(
\frac{1}{|E_\gamma''(0)|} + \frac{1}{|E_\gamma''(\pi)|} 
\nonumber \\
&& +
\frac{2\cos\{t[E_\gamma(0)-E_\gamma(\pi)]+\pi/2\}}{
\sqrt{|E_\gamma''(0)E_\gamma''(\pi)|}} \Bigg) \sim t^{-1}. \qquad
\label{pi_spa}
\eea
The results for the infinite system and arbitrary $\gamma>2$ are readily
obtained: for $\theta_0=0$ we have $E_\gamma(0)=0$  for all $\gamma$ and
$E_\gamma''(0)=2\zeta(\gamma-2)$, where $\zeta(\gamma) \equiv
\sum_{R=1}^\infty R^{-\gamma}$ is the Riemann zeta function, Eq.~23.2.1 of
\cite{abramowitz}. For $\theta_0=\pi$ we get $E_\gamma(\pi) =
E_{\gamma,{\rm max}}$ and
$E_\gamma''(\pi)=2\eta(\gamma-2)=(2-2^{4-\gamma})\zeta(\gamma-2)$, where
$\eta(\gamma) \equiv \sum_{R=1}^\infty (-1)^{R-1} R^{-\gamma}$,
Eq.~23.2.19 of \cite{abramowitz}.  Hence
\bea
\overline{\pi}_\gamma(t) &\approx& \frac{1}{4\pi t} \Bigg\{
\frac{1}{|\zeta(\gamma-2)|} + \frac{1}{|\eta(\gamma-2)|}
\nonumber \\
&& 
-
\frac{2\cos[tE_\gamma(\pi)+\pi/2]}{ 
\sqrt{|\zeta(\gamma-2)\eta(\gamma-2)|}} \Bigg\}.   
\label{pi_spa_inf}
\eea
For $\gamma=3$, this yields $\overline{\pi}_3(t) \approx [2\pi\ln(2)
t]^{-1}$, which also does not show any oscillations.  Comparing
Eq.~(\ref{pi_spa_inf}) for $\gamma=\infty$ to the long-time of the exact
solution \cite{mb2006b}, we have $\overline{\pi}_\infty(t) \approx [2 -
2\cos(4t+\pi/2)]/(2\pi t) = \sin^2(2t+\pi/4)/(\pi t)$, which is exactly
the asymptotic expansion of $\overline{\pi}_\infty(t) = |J_0(2t)|^2
\approx \sin^2(2t+\pi/4)/(\pi t)$, where $J_m(2t)$ is the Bessel function
of the first kind \cite{mb2006b,abramowitz}. 

\begin{figure}[htb]
\centerline{\includegraphics[clip=,width=\columnwidth]{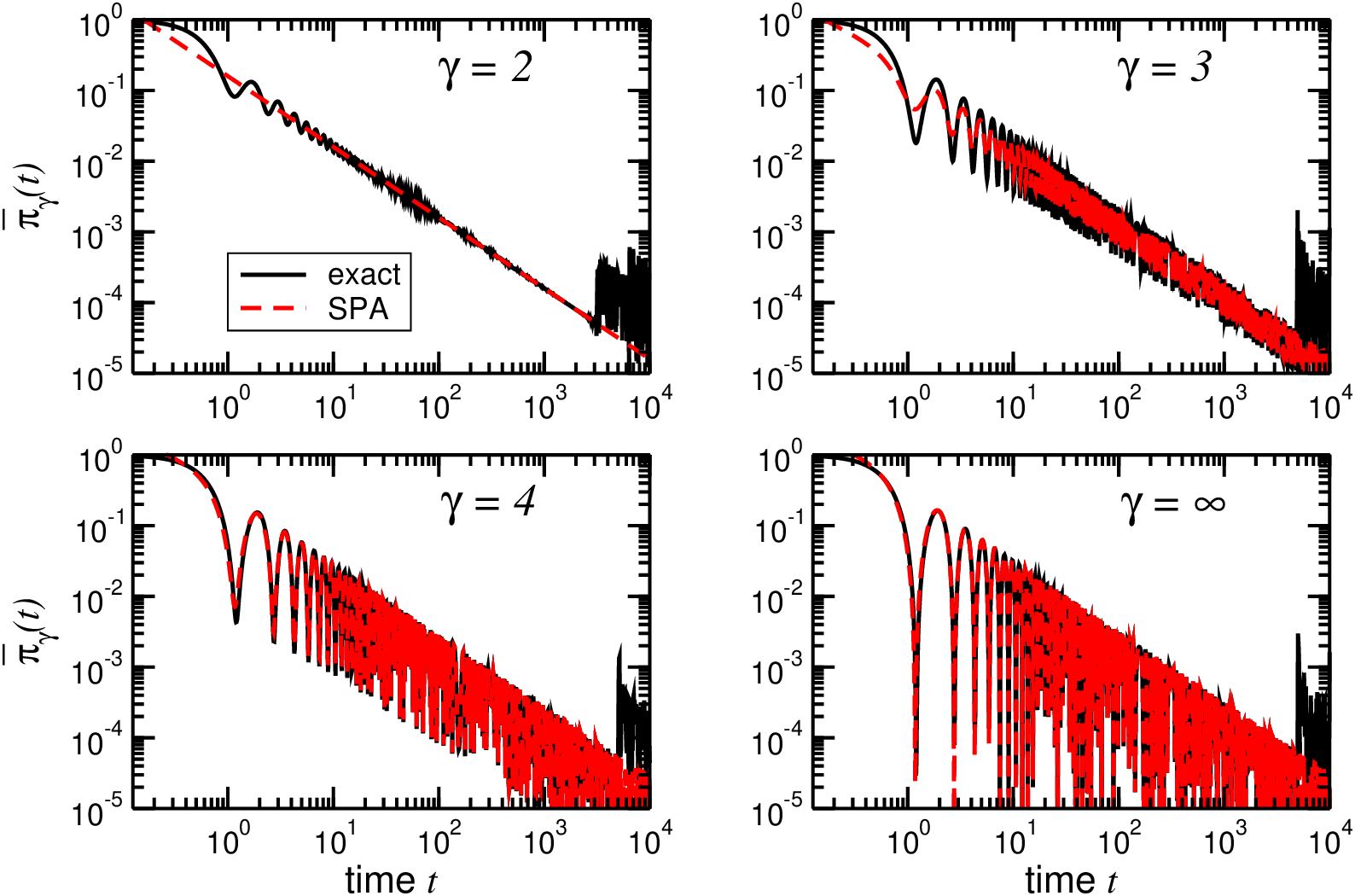}}
\caption{(Color online) Comparison of $\overline{\pi}_\gamma(t)$ obtained
from exact diagonalization (solid black curves) and the SPA (dashed red
curves) for $\gamma=2$, $3$, $4$, and $\infty$.
}
\label{returnprobs_spa}
\end{figure}

Figure~\ref{returnprobs_spa} shows comparisons of
$\overline{\pi}_\gamma(t)$ obtained from the exact diagonalization of
${\bf H}_\gamma$ (solid black curves) to the SPA [Eq.~(\ref{pi_spa}),
dashed red curves]. Clearly, the oscillations decrease with decreasing
$\gamma$. For $\gamma=2$ and large $t$, the oscillations of the exact
$\overline{\pi}_2(t)$ have practically vanished and $\overline{\pi}_2(t)
\approx (2\pi t)^{-1}$. The SPA for $\gamma=3$ still shows oscillations
because $E_\gamma''(0) < \infty$. Increasing $\gamma$ further leads to an
even better agreement of the SPA with the numerically evaluated decay.


In conclusion, we have analyzed the quantum dynamics of excitations on
discrete rings under long-range step lengths, distributed according to
$R^{-\gamma}$. For specific cases, we calculated the DOS analytically and
interpolated to arbitrary step length ranges. The analytically obtained
DOS enabled us to analytically calculate the average probability to be at
the initial site at $t$, which we related to the mean square displacement
at time $t$. The classical MSD show that only CTRW with $\gamma>3$ belong
to the same universality class, displaying normal diffusion. In contrast,
the quantal MSD increase as $t^2$ for {\sl all} extensive cases,
$\gamma\geq2$. Analytic calculations of the probability to be at the
initial node within the stationary phase approximations confirm these
findings. 



Support from the Deutsche For\-schungs\-ge\-mein\-schaft (DFG) and the
Fonds der Chemischen Industrie is gratefully acknowledged.

\end{document}